\documentclass[twocolumn]{aastex701}

\accepted{November 25, 2026}
\submitjournal{PASP}
\begin{document}

\title{The Fast Transient Foreground Fog of LSST}

\author[0000-0001-6013-1131,gname='Guillem',sname='Megias Homar']{G. Megias Homar}
\affiliation{California Institute of Technology, 1200 E. California Blvd., Pasadena, CA 91125, USA}
\affiliation{NSF-DOE Vera C.\ Rubin Observatory, 950 N. Cherry Ave., Tucson, AZ  85719, USA}
\email[show]{gmegias@caltech.edu}
\correspondingauthor{Guillem Megias Homar}

\author[orcid=0000-0001-5390-8563,sname='Kulkarni']{S. R. Kulkarni}
\affiliation{California Institute of Technology, 1200 E. California Blvd., Pasadena, CA 91125, USA}
\email{srk@astro.caltech.edu}  

\author[0000-0002-9601-345X,gname='Craig S.',sname='Lage']{C.~S.~Lage}
\affiliation{Physics Department, University of California, One Shields Avenue, Davis, CA 95616, USA}
\affiliation{NSF-DOE Vera C.\ Rubin Observatory, 950 N. Cherry Ave., Tucson, AZ  85719, USA}
\email{cslage@ucdavis.edu}

\author[0000-0002-9242-8797,gname='J. Anthony',sname='Tyson']{J.~A.~Tyson}
\affiliation{Physics Department, University of California, One Shields Avenue, Davis, CA 95616, USA}
\affiliation{NSF-DOE Vera C.\ Rubin Observatory, 950 N. Cherry Ave., Tucson, AZ  85719, USA}
\email{tyson@physics.ucdavis.edu}

\author[0000-0003-4833-9137,gname='Steven M.',sname='Kahn']{S.~M.~Kahn}
\affiliation{Physics Department,  University of California, 366 Physics North, MC 7300 Berkeley, CA 94720, USA}
\affiliation{NSF-DOE Vera C.\ Rubin Observatory, 950 N. Cherry Ave., Tucson, AZ  85719, USA}
\email{stevkahn@berkeley.edu}

\author[orcid=0000-0003-2992-8024,sname='Harrison']{F. A. Harrison}
\affiliation{California Institute of Technology, 1200 E. California Blvd., Pasadena, CA 91125, USA}
\email{fiona@srl.caltech.edu}

\begin{abstract}
Studies primarily with 1-m class telescopes have now established that 
glints from satellites and space debris are the main obstacles to searches for sub-minute optical transients. Here, we report
a study of this foreground using observations from the Legacy Survey of Space and Time (LSST)
with the Simonyi 8.4-m telescope at the NSF-DOE Vera C. Rubin Observatory.
From 2,955 pairs of consecutive 30-s exposures, spanning 4,755\,deg$^2$ of unique sky area and a total area-time coverage of 472\,deg$^2$\,hr, we identified a population of faint, PSF-like, sub-minute events, including isolated events with no linear sequences of repeated flashes (tracklets) indicative of a tumbling satellite.
We extended the measurements of this population to the faintest magnitudes yet probed, finding a lower-bound to the occurrence rate of $(9.55 \pm 0.14) \times10^6\,\mathrm{sky}^{-1}\,\mathrm{day}^{-1}$ to $\sim23$\,mag (median of $\sim22.3$\,mag).
The large aperture of the Simonyi 8.4-m telescope provides an additional discriminator: orbital debris in low-Earth orbit are sufficiently out of focus that they are filtered out. 
We find that, for a $1^{\prime\prime}$ PSF, unresolved events are constrained to distances beyond $1,732$\,km, and identify a previously unreported population of repeated glints consistent with medium-Earth-orbit altitudes with estimated sizes of order $\sim1-3\,\mathrm{mm}$ and durations of order $\sim10\,{\rm ms}$. We also report two isolated events observed along sight lines deep within the Earth's shadow, for which a sunlit-debris origin is disfavored. The scant evidence is consistent with an eclipsing binary and a stellar flare.
The robust way forward for sub-minute transients is simultaneous observations of the same field by two (or more) widely separated telescopes. The resulting measurement of parallax will 
cleanly eliminate
local (primarily orbital) debris, clarifying the nature of this transient foreground. This understanding will help astronomers to probe truly celestial optical bursts.
\end{abstract}


\section{Introduction}\label{sec:intro}
The search for sub-minute optical bursts (SMOBs) is a frontier in time-domain astronomy.
Such observations may reveal optical counterparts to GRBs and FRBs, sub-minute stellar flares, and potentially new classes of astrophysical phenomena.
Searches in this regime are complicated by a rich foreground of events that mimic astrophysical transients. The list includes
satellite and space-debris glints, detector artifacts, and Solar System objects.

\citet{Schaefer1987} first recognized that satellite glints could masquerade as short-duration astrophysical transients.  The problem has become increasingly important as the population of satellites and orbital debris has grown \citep{Blake_2021, Blake2026, ESA2026}.
\cite{tanaka2025} provides an excellent summary, from the perspective of astronomers,
of the deleterious impact of orbital debris on searches for fast optical
transients from celestial sources. 

The advent of CMOS detectors has enabled searches to explore the
sky on second and sub-second timescales, primarily with
1-m class telescopes (e.g., Evryscope \citep{Ratzloff2019}; W-FAST \citep{Nir2021a}; Tomo-e Gozen \citep{Sako2018}).
A major finding from these searches has been the discovery of glints that, in many instances, appear to be PSF-like
\citep[e.g.,][]{Corbett2020, Nir2021, Karpov2023}.
\setcounter{footnote}{0}
This is what makes them pernicious, since celestial\footnote{We reserve this term for events or objects beyond the solar system.} transients like SMOBs are expected to have a PSF-like
appearance. Separately, 
millisecond-duration Fast Optical Bursts (FOBs), due to their short duration relative to atmospheric seeing, would produce PSFs modulated by the
speckles arising from atmospheric seeing 
\citep{megiashomar2023}.

High-cadence surveys have shown that many such glints last $\lesssim0.5$\,s \citep{Karpov2016, Nir2021}. Some of them, but not all, repeat with short periodicity.
Debris in low-Earth orbits (LEO) have large angular speeds, and for integrating imaging
detectors, 
even short-duration glints result in streaks.
However, more distant orbital populations can remain unresolved.
The interested reader is referred
to Figure~9 of \cite{tanaka2025} for a summary of the glint rate.

The glint rate peaks near the celestial equator, falling 
for $\vert\delta\vert\gtrsim 20^\circ$ \citep{tanaka2025}. The thick equatorial band concentration is due to both active geo-synchronous satellites and older satellites pushed to the geosynchronous Earth orbit (GEO) ``graveyard''.
However, small debris with
large surface-area-to-mass ratios (such as insulation layers)
experience strong solar radiation
pressure and can achieve high orbital inclinations \citep{Pardini2008}.

The glints can be minimized by  looking  
into the shadow cone (umbra) of the Earth. From such observations
\cite{Nir2021} place 
an upper limit of ${\rm 0.052\,deg^{-2}\,hr^{-1}}$ for objects brighter
than 11th mag in Gaia $G_{\rm BP}$. 
Using Tomo-e Gozen, \cite{Arima2026} conducted an intensive search in the deep
umbra of the Earth for fast optical
transients. They found a 17-mag event in Gaia $G$ with a duration of 2\,s and an inferred
rate of $3.4\times 10^{-2}\,{\rm deg^{-2}\,day^{-1}}$.
Excluding this object, the authors placed an upper limit 
on the all-sky rate of $4.17\times 10^{-3}\,{\rm deg^{-2}\,hr^{-1}}$. 
We associate a formal magnitude limit of 17\,mag with this rate.

The ``Deeper Wider Faster" (DWF) survey to identify sub-minute transients was carried out with DECam
on the 4-m Blanco telescope using conventional CCDs with a  20-s 
integration time \citep{andreoni2020}. 
Many M dwarf flares were detected over successive runs. However, last year, 
\cite{Goode2025} identified two candidates of durations of $<20\,$s with point-like PSFs and without quiescent M-dwarf counterparts.

The Legacy Survey of Space and Time \citep[LSST;][]{Ivezic2019} on the 8.4-m Simonyi Survey telescope can extend studies of the glint foreground to substantially fainter events. 
Its effective aperture of 6.49-m enables a 30-s limiting magnitude of $\sim24\,{\rm mag}$. For a 0.5-s glint diluted over a 30-s exposure, the corresponding limiting magnitude is $\sim19.5\,{\rm mag}$, compared with $\leq 18\,{\rm mag}$ reached by Tomo-e Gozen on the Kiso 41-inch in a 0.5-s exposure \citep{tanaka2025}. 
This advantage increases for longer-duration glints. Perhaps, more importantly, its
large aperture also provides a unique advantage: objects in LEO
are out of focus in LSST images, and thus their detection is suppressed \citep{tyson2024}. Conversely, a glint appearing as an in-focus point source must arise above LEO.

Here, we report a study of the glint foreground using LSST observations obtained with a simple setup: pairs of consecutive 30-s exposures of the same field. Such pairs provide a straightforward and robust means of identifying sub-minute events through sources detected in one exposure but absent in the other.

The paper is organized as follows. In \S\ref{sec:data}, we describe the data set and the transient-selection procedure. In \S\ref{sec:results}, we present representative examples, population properties, and occurrence rates of glints in the foreground. In \S\ref{sec:discussion}, we examine events detected within Earth's umbra in our data set, and discuss other recently reported short-duration optical transients. We conclude in \S\ref{sec:conclusion} with the limitations of single-station imaging to distinguish foreground events from astrophysical transients.

We have relegated catalogs and specific technical discussion to the appendix.
The probability theory for point source detection is summarized in
\S\ref{app:chance_detections}.
In \S\ref{app:Defocus} we explain  the focus advantage of the Simonyi 8.4-m telescope.
A brief introduction
to the radial distribution of debris from satellites is given
in \S\ref{app:Satellite_Orbits}.
In \S\ref{app:isolated_catalog}, we present the complete catalog of individual PSF-like events not associated with cataloged satellites or with linear sequences of repeated flashes (``tracklets''), which are characteristic of tumbling satellites.
Finally, in \S\ref{app:tracklet}, we derive orbital-distance and size constraints for the tracklets and work out one representative example.

\section{Observations and Methods}\label{sec:data}
\subsection{Dataset}
We used a data set obtained during commissioning and routine operations of the NSF-DOE Vera C. Rubin Observatory between May 30, 2025 and January 11, 2026. These exposures were not part of LSST science data acquisition, but were taken at the beginning of the night to assess telescope alignment and readiness for science observations. During this procedure, the telescope acquires a series of consecutive, paired 30-s exposures of the same field, with no telescope corrections between exposures. This observing mode resembles the ``snap'' strategy originally envisioned for LSST, in which a visit consisted of two back-to-back 15-s exposures\footnote{Operational changes aimed at improving observing efficiency may soon bring this data-taking mode to an end.}.

\begin{figure}
    \centering
    \includegraphics[width=1\linewidth]{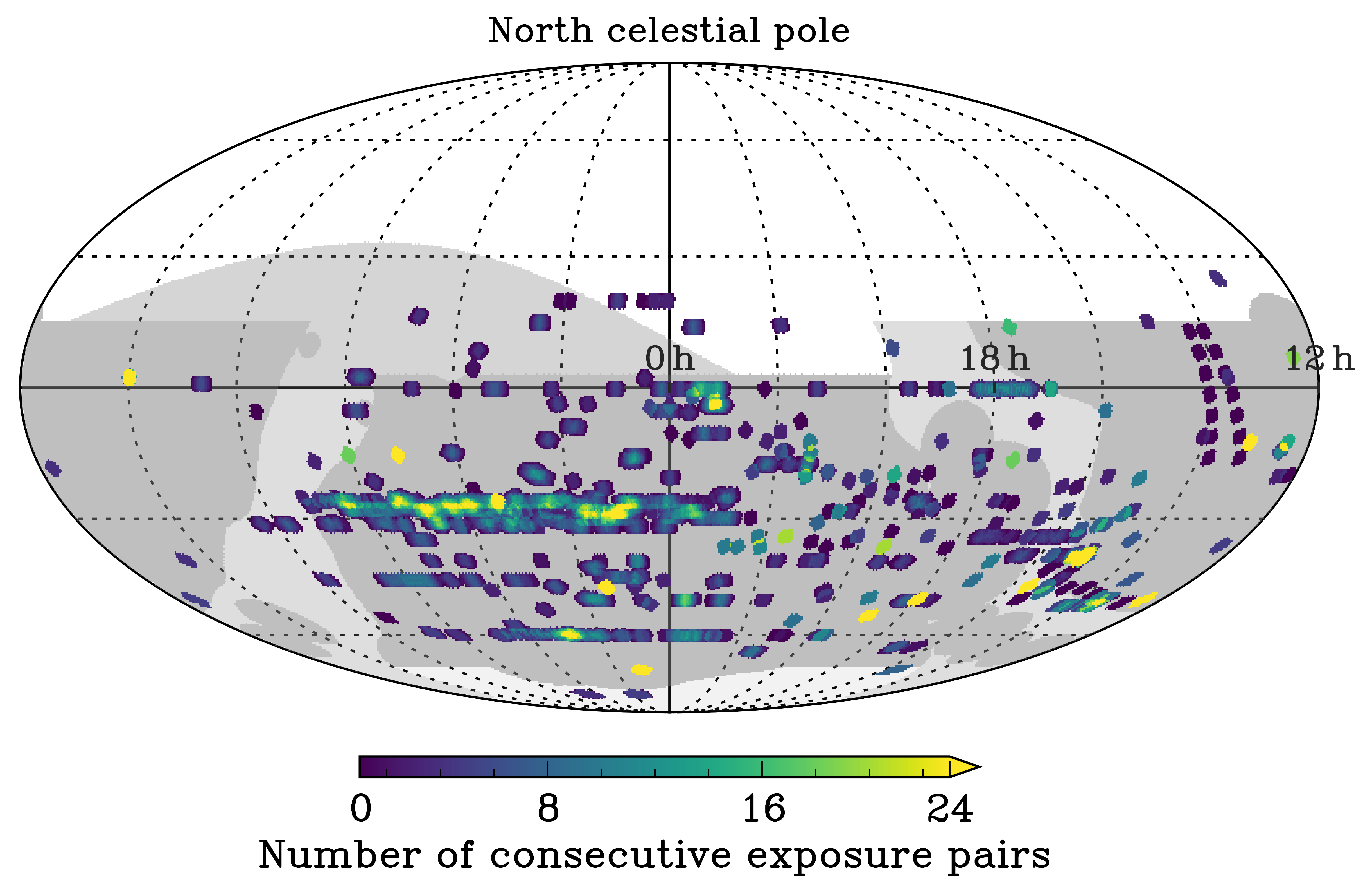}
    \caption{Sky coverage of the LSST consecutive-exposure data set in equatorial coordinates, spanning a total unique area of 4,755\,deg$^2$. The non-uniform coverage reflects the serendipitous nature of the engineering data set, which was acquired during commissioning and routine observatory operations. The gray background indicates the expected 10-year LSST footprint.}
    \label{fig:sky_coverage}
\end{figure}

\begin{table}
\centering
\caption{Summary of the consecutive-exposure dataset by filter. The fraction gives the percentage of the 2,955 exposure pairs obtained in each band.}
\label{table:survey_details}
\begin{tabular}{lrr@{\hspace{-30pt}}r}
\hline
\hline
Band & Pairs & Fraction (\%) &
\begin{tabular}{rrr@{\hspace{0pt}}r}
Total open-shutter\\
time (hr)
\end{tabular} \\
\hline
\hline
$u$ &   26 &  0.9 &  0.43 \\
$g$ &  218 &  7.4 &  3.60 \\
$r$ & 1195 & 40.4 & 19.92 \\
$i$ & 1339 & 45.3 & 22.32 \\
$z$ &  168 &  5.7 &  2.80 \\
$y$ &    9 &  0.3 &  0.15 \\
\hline
Total & 2955 & 100.00 & 49.22 \\
\hline
\hline
\end{tabular}
\end{table}

Our data set comprises 5,910 exposures (2,955 pairs), obtained over 111 nights. 
The selected observations have a median PSF FWHM of $1\farcs07$ and a median ellipticity of 0.098. 
We rejected image pairs with seeing worse than $1\farcs4$ or a telescope-optics contribution exceeding $0\farcs8$, 
as well as exposures affected by substantial image motion during integration.

The sky coverage of these observations is shown in Figure~\ref{fig:sky_coverage}, while their distribution among filters is summarized in Table~\ref{table:survey_details}. Each exposure covers 9.6\,deg$^2$, and the observations collectively span 4,755\,deg$^2$ of a unique sky area. The total open-shutter time is 49.22\,hr, corresponding to an area--time coverage of approximately 472\,deg$^2$\,hr. Across the data set, the full LSST field lies within the earth's umbra for 1,905 pairs (64.4\%) at 1,000\,km, 204 pairs (6.9\%) at 10,000\,km, and only 7 pairs (0.2\%) at GEO altitudes, after accounting for $70^\prime$ of atmospheric refraction in determining the umbra boundary.

\subsection{Data Reduction} 
All images are processed through standard LSST calibration pipelines \citep{pipelines2026}, including instrument signature removal, detector-artifact repair (including cosmic rays, crosstalk, and alpha-particle events), photometric, and astrometric calibration. For each pair of consecutive exposures, the second visit is warped onto the WCS of the first and the PSF matched to it using the Alard--Lupton algorithm \citep{AlardLupton1998}, with the first visit serving as the science image. The resulting difference image therefore retains the PSF of the first visit.
The
initial list of transients is obtained with 
a threshold for source detection in the difference
image set at $5\sigma$.
Detections falling on flagged pixels 
({\it i.e.}, saturated, cosmic-ray, interpolated, suspect, bad, edge, no-data, cross-talk or unmasked) 
in either visit are rejected. 

To suppress chance noise detections, we further restrict the sample to detections with ${\rm SNR}>10$ in the difference image (see \S\ref{app:chance_detections} for further discussion). This also ensures that the source morphology is measured with sufficient precision.

Next, we reject detections inconsistent with genuine flux variability, which are predominantly poor subtraction residuals.  For each candidate, we perform forced PSF photometry at the same position in both visits and define a statistic, 
\begin{equation}
r = \frac{|\Delta f|}{\max(f_1, f_2)},
\end{equation}
where $f_1$ and $f_2$ are the measured fluxes. A transient present in only one exposure has $r \simeq 1$, whereas a source with comparable flux in both exposures has $r \ll 1$. We retain detections with $0.5 \le r \le 1.5$. The lower bound corresponds to a flux change of at least a factor of two. The upper bound allows the forced flux of a source absent in one exposure to fluctuate below zero because of background noise, while rejecting large negative excursions. The observed distribution of the final sample peaks at $r=1$, with a broader tail toward lower $r$, where residual flux present in both visits is expected to contribute.

Trailed sources are identified by trailed source measurement flags and excluded. Additionally, sources for which their adaptive Gaussian-weighted central moments \citep[HSM;][]{Hirata2003, Mandelbaum2005} cannot be measured are also excluded, resulting in the exclusion of a large subset of sources that are trailed.

Tumbling satellites are identified by searching for 
tracklets consisting of at least three detections aligned along a linear trajectory, using a Random Sample Consensus (\texttt{RANSAC}) line-fitting algorithm \citep{fischler1987}.
This search is performed both within individual detectors and across the full focal plane at a given epoch, including all detections with $\mathrm{SNR}>5$ to prevent $\mathrm{SNR}>10$ sources from being misclassified as isolated when they belong to tracklets with fainter members.
The focal-plane search is restricted to tracks that span more than $0.3^\circ$.
The sources are then cross-matched against the NORAD satellite catalog using two-line elements obtained from Space-Track\footnote{https://www.space-track.org/, operated by U.S. Space Force.} propagated to the exposure midpoint with SGP4 \citep{Vallado2006}.
A detection is matched when it
lies within $0.05^\circ$ of the predicted position perpendicular to the apparent track and $0.5^\circ$ along it, the asymmetry reflecting the along-track dominance of SGP4 error. Detections associated with tracklets or known satellites are flagged but retained in the sample.

\begin{table}
\centering
\caption{Summary of the successive selection criteria applied to the difference-image detections. The fraction is relative to the initial sample.}
\label{tab:dataset_cuts}
\begin{tabular}{lrr}
\toprule
Stage & Detections & Fraction (\%)\\
\hline
\hline
Initial sample                & 33,816,942 & 100.000 \\
$\mathrm{SNR}\geq10$                          & 3,940,776    & 11.653 \\
Forced-photometry consistency                        & 112,009  & 0.331   \\
Removal of trailed detections                              & 18,731    & 0.055   \\

\hline
\multicolumn{3}{l}{\textit{Morphological selection:}} \\
\quad Measurable HSM shape                  & 8,921    & 0.026 \\
\quad PSF similarity, $\rho_{\mathrm{PSF}}$ & 7,321    & 0.022 \\
\quad Minor-axis ratio,
  $\sigma_{\min}^{\rm src}/\sigma_{\min}^{\rm PSF}$ & 5,325	 & 0.016 \\
\quad Trace ratio,
  $T_{\rm src}/T_{\rm PSF}$                 & 5,087    & 0.015 \\
\quad Broadness, $\kappa\sigma^4$           & 4,901    & 0.014 \\
\hline
\hline
\end{tabular}
\end{table}

For each remaining detection, a $5\times5$ arcsec difference-image cutout is compared with the PSF model evaluated at the detection position. We quantify their morphological agreement using the PSF similarity,
\begin{equation}
    \rho_{\mathrm{PSF}} = \max_{\Delta x, \Delta y} \frac{\left|\Sigma_i D_i P_i(\Delta x, \Delta y)\right|}{(\Sigma_i D_i^2)^{1/2} [\Sigma_i P_i^2(\Delta x, \Delta y)]^{1/2}}
\end{equation}
where $D_i$ and $P_i$ are the pixel values of the difference image and the PSF model, respectively, and the maximization over translations of the PSF model makes the statistic sensitive to morphology rather than centroid errors. $\rho_{\mathrm{PSF}}$ approaches unity when the PSF model reproduces the stamp exactly and decreases as the morphology departs from that of a point source. Dipole residuals, for example, are antisymmetric and cannot be modeled by the PSF, yielding $\rho_{\mathrm{PSF}}$ near zero. At low SNR, noise can drive $\rho_{\mathrm{PSF}}$ to low values even for point sources. At ${\rm SNR}>10$, the source signal dominates sufficiently that we can reject detections with $\rho_{\mathrm{PSF}}<0.5$.

To reject artifacts for which $\rho_{\mathrm{PSF}}$ is less informative, we apply three complementary morphological cuts: 
(1) the PSF trace ratio, $T_{\mathrm{src}}/T_{\mathrm{PSF}}$,
where $T=I_{xx}+I_{yy}$ is the trace of the adaptive second-moment matrix,
which removes compact cosmic-ray events; 
(2) the minor-axis ratio, $\sigma_{\mathrm{min}}^{\mathrm{src}}/\sigma_{\mathrm{min}}^{\mathrm{PSF}}$, 
where $\sigma_{\min}^2 = T/2 - \sqrt{(T/2)^2 - (I_{xx}I_{yy}-I_{xy}^2)}$,  
which rejects features narrower than the PSF along either axis; and 
(3) the broadness statistic $\kappa\sigma^4$, 
the weighted fourth radial moment normalised by the squared second moment \citep[see][]{MegiasHomar2026}. 
We evaluate it with a fixed Gaussian weight of width $1.5\,\sigma_{\mathrm{PSF}}$
rather than the adaptive weight, for both the source and the PSF model at the
same position.
This rejects alpha-particle events in the camera that can match the PSF in second-moment width while remaining more centrally concentrated in their radial profile \citep{MegiasHomar2026}.

For each statistic, we reject detections whose ratio falls below the first percentile of the corresponding simulated point-source distribution at the same SNR.
These distributions are constructed from 160,000 simulated point sources spanning 16 SNR values, using 250 representative PSF models with 40 independent noise realizations each.
Thus, each threshold is designed to retain 99\% of genuine point sources at a given SNR, while rejecting sources with anomalously low values of the corresponding statistic.
The cuts are one-sided, rejecting only sources narrower or more centrally concentrated than the PSF, since convolution with the PSF cannot produce a source narrower than the PSF itself.

A summary of the resulting candidate population and the effect of each selection criterion are provided in Table~\ref{tab:dataset_cuts}.

\begin{figure*}[ht!]
    \centering
    \includegraphics[width=0.9\linewidth]{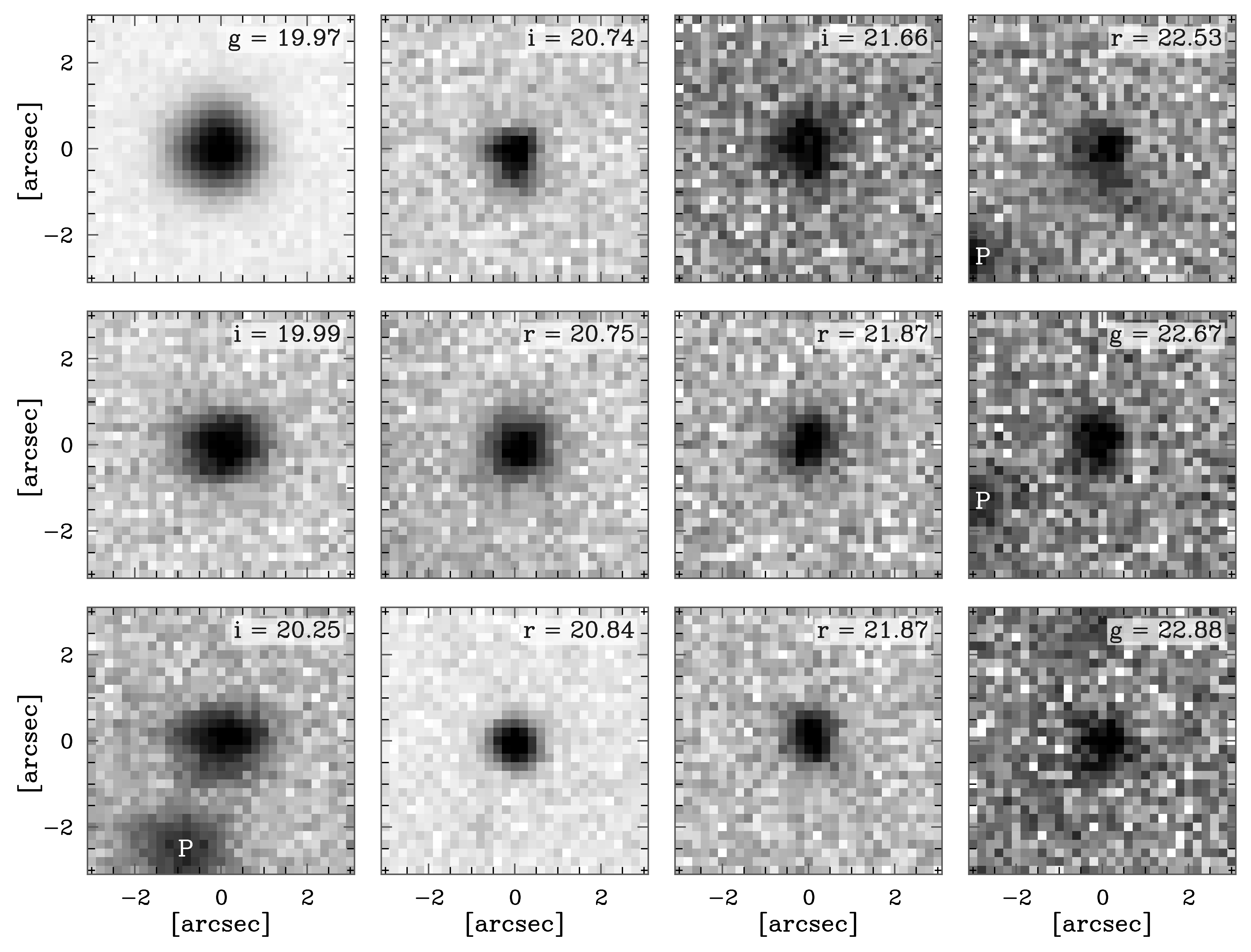}
    \caption{Selected cutouts from the 424 isolated, PSF-like transient foreground events, ordered by brightness from the brightest at the top left to the faintest at the bottom right. The transient event is centered in each panel; additional faint sources visible nearby are persistent sources in the underlying field (marked with a ``P'').  Most are likely satellite or space debris glints, despite appearing isolated rather than as tracklets and lacking matches to known cataloged objects. Some exhibit structured morphologies, suggesting that more complex object or reflection geometries may be required to explain their appearance.}
    \label{fig:science_cutouts}
\end{figure*}

\begin{figure*}[ht!]
    \centering
    \includegraphics[width=\linewidth]{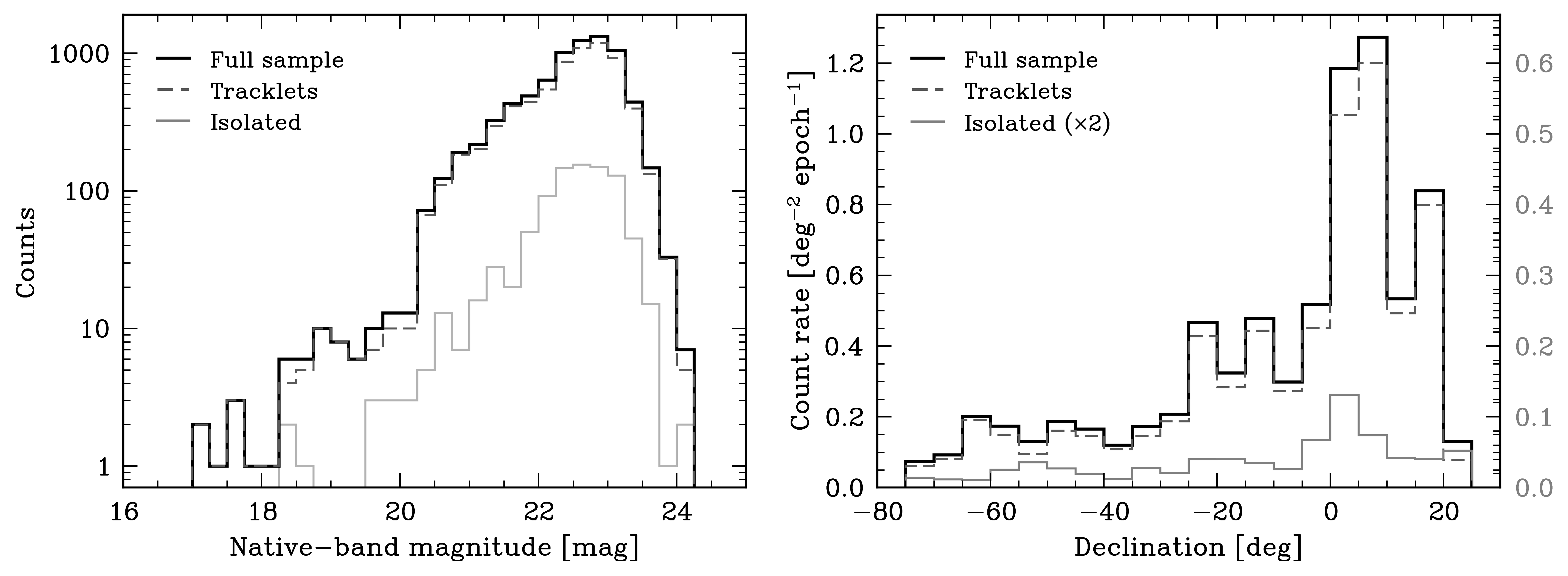}
    \caption{Magnitude distribution and declination-dependent occurrence rate of the sub-minute foreground transients. (\textit{Left}) Magnitude distributions in the native LSST band of each detection on the AB system. The black curve shows the full sample, while the dashed and gray curves show the satellite/tracklet-associated (90.63\%) and isolated (8.7\%) populations, respectively. (\textit{Right}) Occurrence rate as a function of declination, normalized by the survey effort in each declination bin. The isolated-sample curve is multiplied by two; the right-hand vertical axis gives its unscaled rate.}
    \label{fig:mag_hist}
\end{figure*}

\section{Results}\label{sec:results}

\subsection{Candidate Population}
Our search yields 4,901 PSF-like events with $\mathrm{SNR}>10$. Of these, 90.63\% are associated with identified tracklets containing three or more collinear glints spanning at least $0.3^\circ$ of the focal plane, while an additional 0.7\% lie on shorter collinear sequences confined to a single detector. This leaves 424 (8.7\%) fully isolated PSF-like detections without such associations.

Incidentally, if the minimum multiplicity of tracklets is increased from three to five collinear glints, the isolated sample increases to 1,077 events (22.38\%). Throughout the paper, we adopt the more conservative three-glint criterion to minimize contamination from unidentified tracklets. We cannot rule out the association of pairs of isolated events, as any two points can trace a line.
In addition, because trailed sources are excluded from the sample (see \S\ref{sec:data}), our tracklet search may be incomplete, as trailed sources could align with unresolved glints along the same trajectory.

Figure~\ref{fig:science_cutouts} shows representative examples drawn from the sample of 424 isolated events.
The PSF-like morphology of these events implies that,
if they arise from artificial objects,
the objects must lie beyond LEO,
as otherwise they would appear out of focus
due to the large aperture of the Simonyi 8.4-m telescope (see \S\ref{app:Defocus}).
This means that the events are either in Medium Earth Orbit (MEO) or
GEO or celestial (see \S\ref{app:Satellite_Orbits}
for a brief summary of orbits of earth-bound debris).

Galactic events should have a 
stellar counterpart (e.g., flare stars), unless they belong to a previously unrecognized population, whereas an extragalactic origin would favor an associated host galaxy at or near the event position. To test the latter, in \S\ref{app:isolated_catalog},
for the isolated sample,
we report the position and magnitude of each event together with the nearest cataloged galaxy and the corresponding probability of chance coincidence. 
For 73\% of the isolated events, the closest-galaxy associations are statistically consistent with chance projection ($P_{\rm chance} > 0.1)$, 
further supporting the interpretation that 
the isolated population is dominated by foreground events. However, a minority of the sample may be astrophysical, as illustrated by the two events detected within the Earth's umbra (see \S\ref{sec:umbra}). 

Across the full sample, only 39 events coincide with objects in the NORAD catalogue.
This low match rate is unsurprising given the catalogue's orbital distribution: of its 31,048 objects, 26,236 (84.5\%) are in LEO and only 572 (1.8\%) in MEO. Our selection excludes LEO objects (see \S\ref{app:Defocus}), while our results suggest that many of our detections originate at MEO distances (see \S\ref{sec:meo_orbits}).

\subsection{Magnitude and Sky Distribution}
Figure~\ref{fig:mag_hist} summarizes the properties of the resulting foreground population. The observed magnitude distribution has a median of $\sim22.3$\,mag. This distribution combines observations obtained in different LSST filters and, therefore, represents the native-band AB magnitudes of the detections rather than photometry in a common band. Since our data set did not uniformly sample the sky, we compute the relative importance of the number of candidates as a function of declination, which shows a peak of declination angles around $\delta\approx 0^\circ$.

\subsection{Tracklet Altitudes and Medium-Earth Orbit Objects} \label{sec:meo_orbits}
Our search identifies 1,299 unique tracklets containing at least one glint with $\mathrm{SNR}>10$, accounting for 90.63\% of the 4,901 events in our sample.
Because tracklet identification includes all detections with $\mathrm{SNR}>5$, we also identify 93,562 tracklets containing no sources above $\mathrm{SNR}>10$.
However, at the $5\sigma$ threshold, approximately $8\times10^2$ spurious detections are expected in each exposure pair (see \S\ref{app:chance_detections} for more details), which undermines the reliability of these fainter tracklets.

\begin{figure*}
    \centering
    \includegraphics[width=.8\linewidth]{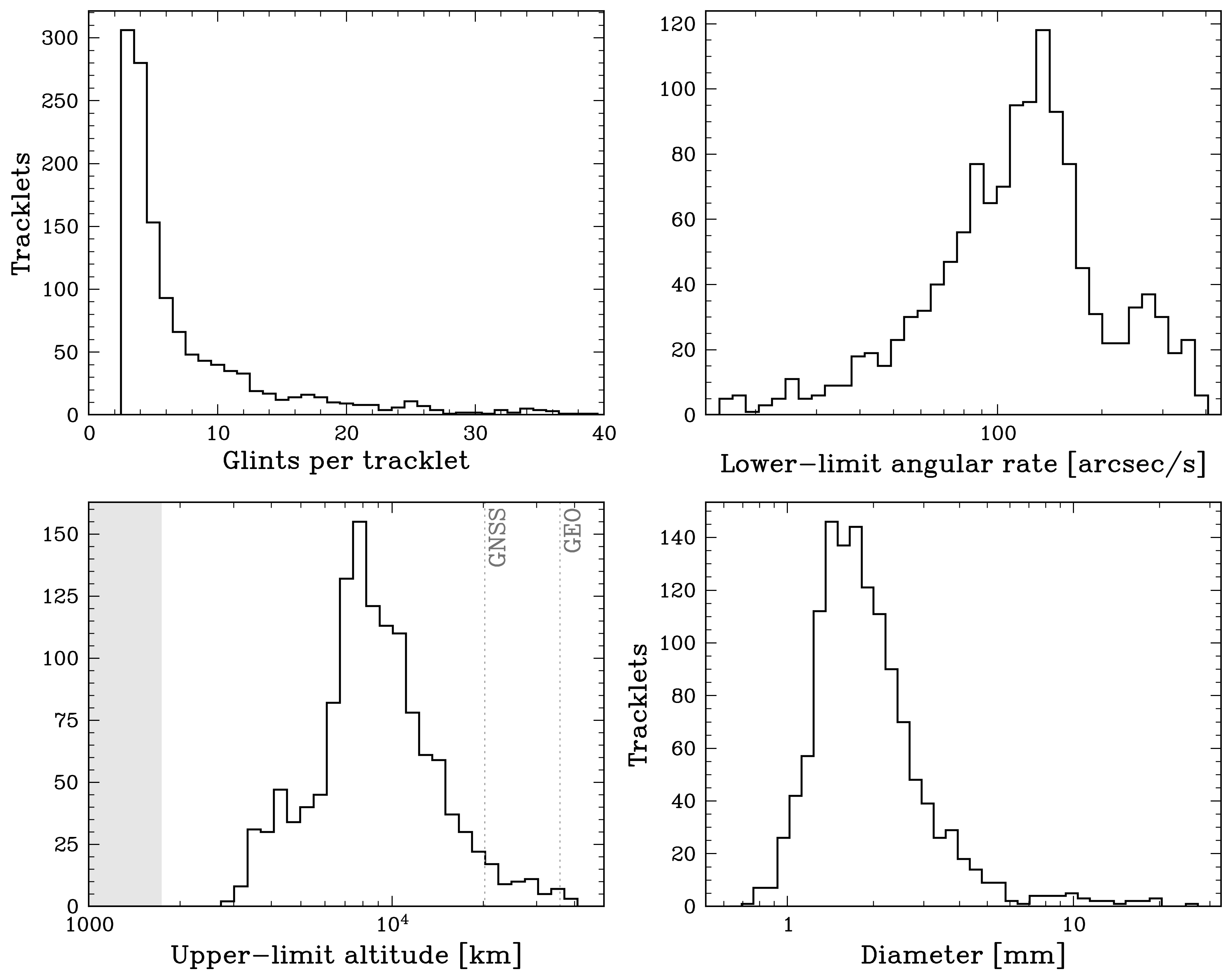}
    \caption{
    (\textit{Top left}) Number of glints per tracklet for tracklets containing at least one detection with $\mathrm{SNR}\,>\,10$, counting all associated glints with $\mathrm{SNR}\,>\,5$. Sixteen tracklets lie beyond the plotted range.
    (\textit{Top right}) Lower limits on the apparent angular rates, inferred from the angular extent of each tracklet and the maximum time interval over which its glints could have occurred (either 30\,s or 60\,s).
    (\textit{Bottom left}) Upper limits on tracklet altitudes above the Earth's surface, inferred from the angular extent of each tracklet on the sky and obtained by solving Eq.~(\ref{eq:altitude}) for $h$. The gray region marks distances at which objects are expected to appear out of focus in LSST images ($<\,$1,732\,km; see \S\ref{app:Defocus}). The distribution peaks near 8,000\,km. (\textit{Bottom right}) Upper limits on reflector sizes assuming perfect specular reflection and an albedo of $\rho=0.175$. Most inferred sizes are $\sim1$--$3\,{\rm mm}$. See \S\ref{app:tracklet} for more details on both calculations.
    }
    \label{fig:tracklet_hist}
\end{figure*}

Figure~\ref{fig:tracklet_hist} shows the number of glints per tracklet, lower limits on angular velocities, upper limits on orbital altitude, and size distributions for the 1,299 unique tracklets.
The top left panel shows that most tracklets contain only three or four glints, highlighting the sensitivity of our results to the minimum requirement of three collinear detections.
But are all these tracklets genuine?
Some may arise from chance alignments of otherwise isolated glints.

The top right panel shows the lower limits on angular velocities inferred from the observed tracklet spans, 
while the bottom left panel shows the corresponding upper limits on orbital altitudes, 
assuming a circular Keplerian orbit (see \S\ref{app:tracklet} for more details).
The slight bimodality in the angular velocity distribution reflects the two time windows used to estimate these lower limits: 30\,s for tracklets confined to a single exposure and 67\,s for those spanning both exposures.
Since a tracklet may extend beyond the field of view, its observed angular span provides only a lower limit on the angular velocity, and hence an upper limit on the orbital altitude.
The resulting distribution reveals a large population of objects at MEO altitudes. 

The bottom right panel shows the inferred reflector-size distribution. 
For each tracklet we take its brightest glint and correct the 30-s magnitude for the flash duration:
since every glint is unresolved, its duration is constrained by the time required for the object to move by one seeing element, $t_{\rm g}\le 1''/\omega$, where $\omega$ is the apparent angular rate. These upper limits on the glint duration have a median of 8\,ms, with a range of 4--19\,ms across the tracklet sample. Assuming perfect specular reflection and an albedo of $\rho=0.175$ \citep{Africano2004}, we then use the inferred instantaneous magnitudes and orbital altitudes to estimate the diameters of the reflecting surfaces (see \S\ref{app:tracklet} for more details). 
The resulting characteristic sizes are of order $1$--$3\,{\rm mm}$.
These estimates represent lower limits at the assumed
orbital altitudes, since shorter glint durations or
less efficient reflection would require larger surfaces.
However, because the inferred orbital altitudes are
themselves upper limits, closer objects could produce
the same brightness with smaller reflecting surfaces.

\subsection{Occurrence Rates}
We identify 4,555 PSF-like glints with $\mathrm{SNR}>10$ down to $\sim23$\,mag.
Given the effective sky--time coverage of our search,
this corresponds to an occurrence rate of $9.65 \pm 0.14\,\mathrm{deg}^{-2}\,\mathrm{hr}^{-1}$, where the uncertainty reflects counting statistics only.
This corresponds to an average, cumulative rate of $(9.55 \pm 0.14) \times10^6\,\mathrm{sky}^{-1}\,\mathrm{day}^{-1}$. We quote the rate to this magnitude because the observed magnitude distribution in Figure \ref{fig:mag_hist} begins to turn over at $\sim23$\,mag, indicating increasing incompleteness at fainter magnitudes.

Our measured rate is of the same order as that reported by \citet{tanaka2025}, who find an average cumulative glint rate of $\sim6\times10^6\,\mathrm{sky}^{-1}\,\mathrm{day}^{-1}$.
Their limiting magnitude of $\sim18$\,mag in 0.5-s exposures corresponds to $\sim22.4$\,mag when diluted over a 30-s exposure, approaching the depth of our search, which extends to $\sim23$\,mag for sub-second events.
The two measurements, however, are not comparable:
although the angular velocities of many glints detected by \citet{tanaka2025} are consistent with GEO objects, 
their selection does not exclude LEO glints, 
whereas our selection filters out LEO objects, which appear defocused due to the large mirror aperture.
Our measured rate therefore represents only the more distant artifical foreground, primarily MEO and GEO.
The total glint rate at LSST depths, including LEO objects, would thus be larger than the rate reported here.
The rate also depends on the observed declinations and on whether the observed fields lie within the Earth's umbra.
Our measurement represents a cumulative rate averaged over the survey observations, rather than a rate representative of the entire sky.

Among the seven pairs of exposures fully within the Earth's umbra beyond $35{,}000$\,km, we identified two PSF-like events (see \S\ref{sec:umbra} and Figure \ref{fig:umbra_events}), which are most likely a stellar flare and an eclipse, respectively. Excluding these events, the 1.12\,deg$^2$\,hr of umbra coverage yields a 95\% Poisson upper limit on the SMOB rate of $2.675\,\mathrm{deg}^{-2}\,\mathrm{hr}^{-1}$ to $\sim24$\,mag. Although weaker than previous constraints \citep{Nir2021, tanaka2025}, this measurement extends the search for SMOBs within the Earth's umbra to fainter magnitudes.

\section{Discussion}\label{sec:discussion}

The measured occurrence rate is similar to that reported by \citet{tanaka2025}, despite important differences between the two surveys. Tomo-e Gozen used 0.5-s exposures and included glints from LEO objects, whereas our 30-s LSST exposures effectively filter out LEO objects through defocus. The large rate measured here therefore indicates that the foreground from MEO and GEO objects alone is substantial. Although short ($\lesssim 0.5$\,s) glints are diluted in LSST exposures, they remain detectable thanks to its large aperture.

Because our selection criteria (\S\ref{sec:data}) are deliberately conservative and no completeness correction is applied, the rates reported here should be regarded as lower limits. For events not associated with identified satellites or tracklets, we do not attempt to assign a definitive origin on an event-by-event basis (we report the full sample in \S\ref{app:isolated_catalog}).

\subsection{Evidence for an Orbital Origin}

Despite the absence of identifiable tracklets for many events, the declination distribution shown in Figure \ref{fig:mag_hist}, with most candidates concentrated between $\vert\delta\vert\lesssim 20^\circ$, points to geosynchronous satellites and space debris as the dominant contributors. However, our tracklet analysis indicates that most tracklets are instead consistent with MEO altitudes, with inferred reflector sizes of order $1$--$3\,{\rm mm}$ and durations of $\sim 10\,{\rm ms}$. This population is unexpected, 
as noted in \S\ref{app:Satellite_Orbits}, as relatively few debris objects are known or predicted to occupy MEO altitudes.

The MEO debris population is poorly constrained.
Public catalogues are considered nearly complete only above $\sim10$\,cm in LEO and $\sim1$\,m in GEO, with no comparable completeness limit for MEO \citep{matney2025}.
NASA's Orbital Debris Engineering Model (ORDEM) and ESA's Meteoroid and Space Debris Terrestrial Environment Reference (MASTER) model are constrained primarily by LEO radar and GEO optical measurements \citep{matney2025,horstmann2025},
while the only dedicated optical survey of MEO detected no uncatalogued objects \citep{schildknecht2012}.
ESA listed 1,303 objects, excluding navigation satellites, in the MEO class at the end of 2025 \citep{ESA2026}. Some of our tracklets may correspond to these objects, but the large number of tracklets detected in our observations (1,299) is difficult to reconcile with the MEO population represented in current public catalogues.

The concentration toward low declinations is less pronounced in the isolated sample, which also contains events at larger declinations. Such events do not necessarily rule out a geosynchronous origin, since solar radiation pressure can drive substantial inclination and eccentricity to small debris with large surface-area-to-mass ratios (see \S\ref{app:Satellite_Orbits}).
Slowly tumbling multi-layer insulation (MLI) shards may naturally produce such isolated events: eddy-current damping can reduce their spin rates, allowing small mirror-like surfaces to generate only occasional glints.

Visual inspection of a randomly selected subset of isolated events (see Figure~\ref{fig:science_cutouts}) reveals another notable feature: some exhibit structured or multi-component PSFs.
Such morphologies may arise from multiple reflecting surfaces on a single object, while others could be consistent with slightly out-of-focus glints. 
Very short-duration flashes may also exhibit structured PSFs because they sample the instantaneous atmospheric turbulence rather than the time-averaged seeing disk \citep{megiashomar2023}.
A wide range of object geometries, reflective configurations and flash durations could plausibly produce these signatures.

\begin{figure*}[ht!]
    \centering
    \includegraphics[width=\linewidth]{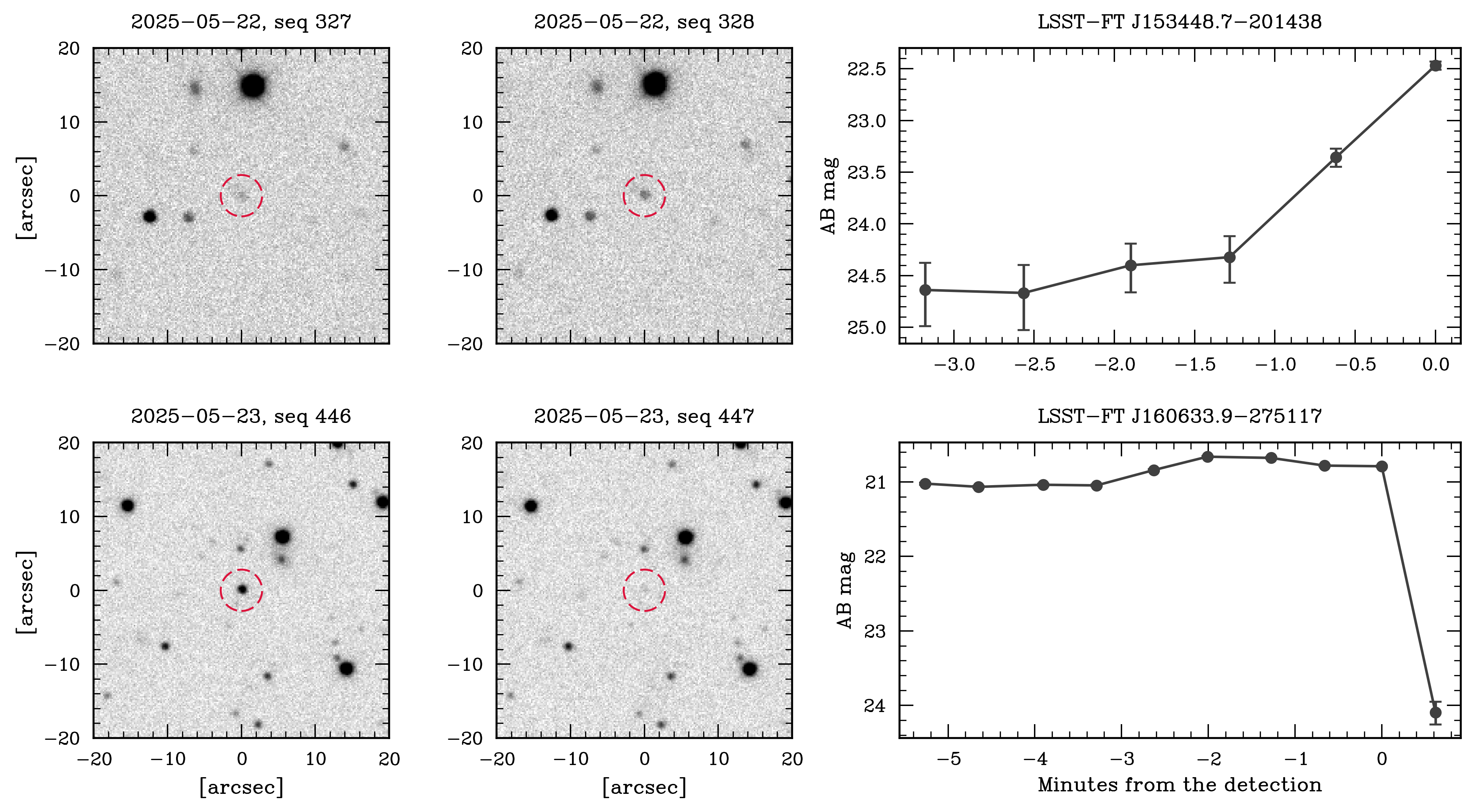}
    \caption{Two transient events detected in observations in $r$-band for which the line of sight lies deep within the Earth's umbra. For the first event, the umbra extends beyond 40,000\,km and no cataloged source is found at the transient position. The second event, LSST-FT J160633.9$-$275117, was observed with an umbra height of 35,018\,km and is likely an eclipsing stellar source. No additional observations are available in this dataset, and follow-up would be required to establish the nature of either event.}
    \label{fig:umbra_events}
\end{figure*}

\subsection{Constraints on Foreground Object Size and Distance}\label{sec:angular}

Figure~\ref{fig:tracklet_hist} shows that most tracklets have altitude upper limits in the MEO regime. None of these upper limits falls below $\sim1,732$\,km, consistent with the LSST defocus limit discussed in \S\ref{app:Defocus} and shown in Figure~\ref{fig:defocus_curve}. We examine a tracklet in detail in \S\ref{app:tracklet}. Its observed angular motion implies a distance of 7,269\,km and orbital speed of $5.4\,\mathrm{km}\,\mathrm{s}^{-1}$, consistent with the distance required for the object to remain unresolved.

The identified tracklets allow us to roughly estimate the spatial density of the detected debris population,
rather than the occurrence rate of individual glints,
which includes multiple flashes from the same object.
Between 2,000 and 20,000\,km, we identify 1,233 tracklets (94.9\% of the 1,299 unique tracklets).
Assuming a field of view of 9.6\,deg$^2$ and accounting for all 2,955 exposure pairs, the total surveyed volume between 2,000--20,000\,km is $2.3\times10^{13}\,\mathrm{km^3}$.
The corresponding debris spatial density is $5.36\times10^{-11}\,\mathrm{km^{-3}}$.

The unresolved morphology of the glints constrains their durations to order $\sim10\,\mathrm{ms}$ (see \S\ref{app:tracklet} for the derivation), substantially shorter than the 0.1--0.5\,s durations previously reported for GEO glints \citep{Karpov2016,Nir2021}.
At such short durations, one would expect the structured PSF signatures discussed by \citet{megiashomar2023}.
These are difficult to identify by eye at the LSST pixel scale of $0\farcs2$ (see, e.g., their Figure 1).
Our simple PSF-based morphological selection is not designed to distinguish this structure.
Several glints in the representative tracklet shown in Figure \ref{fig:tracklet} seem to exhibit an apparent substructure that changes along the tracklet.
A detailed analysis of these morphologies, however, is beyond the scope of this work.

Assuming an albedo of 0.175 \citep{Africano2004} and perfect specular reflection, we estimate characteristic reflector sizes of order 1--3\,mm.
For comparison, a diffuse Lambertian reflector at a phase angle of $90^\circ$ would require sizes larger by a factor of $\sim500$ \citep{tanaka2025}.
However, aluminized multilayer insulation (MLI) is predominantly specular \citep{Fruh2012}.

However, some identified tracklets do not necessarily represent repeated glints of a single tumbling object. Fragmentation or material shed from a satellite can produce populations of debris with slightly different orbital energies and periods, causing individual fragments to drift apart and spread along the parent orbit over time \citep{Johnson2009, Shu2026, Rahman2026}. If several of such fragments become reflective during the same exposure, they could appear as a sequence of approximately collinear, spatially separated point sources, providing an alternative potential origin for some of the observed tracklets. This scenario could explain why the glints in some tracklets appear at nonsequential positions along their trajectories (e.g., see Figure \ref{fig:tracklet})

\subsection{Events in the Earth's Umbra}\label{sec:umbra}
Among the seven pairs of exposures for which the Earth's umbra extends beyond 35,000\,km along the line of sight, only two events pass our criteria. We show these events in Figure~\ref{fig:umbra_events}, together with the corresponding light curves. Both were observed in the $r$-band. They appear to be genuine transient events. However, because these observations were obtained serendipitously, no subsequent imaging in the data set is available to constrain their evolution. A dedicated program using the same consecutive-exposure strategy but continuously monitoring a field within Earth's umbra would provide stronger constraints on the nature of such events.

LSST-FT J153448.7$-$201438 has no counterpart among known sources. The nearest Gaia source lies at a separation of $12\farcs91$. No obvious galaxy is found in the immediate vicinity. The nearest cataloged galaxy lies at a separation of $15\farcs24$. The most probable host candidate lies at a separation of $15\farcs95$, has $r=20.97$\,mag, and $P_{\rm chance}=0.247$. In its position, the 4-m Blanco telescope Legacy Survey DR10 reaches a limiting depth of $g=23.59$\,mag. The rapid rise is, perhaps, reminiscent of a stellar flare, but in the absence of later-time observations or follow-up data, its origin cannot be established.
If a flare star, then 
future deeper $r$-band imaging should reveal the quiescent counterpart to the star.

LSST-FT J160633.9$-$275117 is likely an eclipsing stellar source. Its position coincides with Gaia DR3 6042088459340233856 at a separation of $0\farcs06$, a $G=20.82$\,mag source with proper motion $7.945$\,mas\,yr$^{-1}$ and a parallax of $3.112\pm3.463$\,mas, which is not significant ($0.9\sigma$). One possible interpretation is an eclipsing white-dwarf binary, for which ingress times of $\sim30$\,s are not unusual. However, the observed  drop of $\sim2$\,mag implies
\begin{equation}
    \frac{F_{\rm min}}{F_{\rm out}}=10^{-0.4(2)}=0.158,
\end{equation}
so that, for a total eclipse, the eclipsed component must contribute $\sim84.2\%$ to the out-of-eclipse flux. Together with the relatively red Gaia color $G_{\rm BP}-G_{\rm RP}\simeq1.0$, this would favor a configuration in which a relatively cool white dwarf dominates while the companion is a faint low-mass M dwarf. Spectroscopic or multi-band follow-up would be required to test this interpretation, which we do not attempt to establish here.

\subsection{Other Events}
The results presented here have implications for the interpretation of recently reported fast optical transient candidates.  We note that several of the morphologies shown in Figure~\ref{fig:science_cutouts} and \ref{fig:isolated_cutouts} closely resemble those of the two 20-s events reported by \citet{Goode2025}. Their observations had an umbra height of only about 800\,km, leaving more distant debris illuminated. PSF-like glints are known to arise from objects beyond $\sim1000$\,km, and our results further demonstrate that a substantial population of such events can appear in isolation, without an associated tracklet. Moreover, the event rate reported by \citet{Goode2025} is below the glint rates reported here and by \cite{tanaka2025}. Together, these considerations suggest that the simplest interpretation of their two events is glints from orbital debris.

The 2-second duration event reported by Tomo-e Gozen \citep{Arima2026} was
observed in the deep umbra of the Earth, and the brightness of the event (17th mag)
is a real puzzle, as is the fact that the structure of the source evolves over two seconds!
The authors explore a collision model (solar system objects) and 
atmospheric phenomena (meteors) and find the rates to be so low as to 
render the hypotheses untenable.
Nonetheless, in our view, the change in the object's appearance requires that the event must be local.

\section{Conclusion}\label{sec:conclusion}

We have presented a catalog of faint PSF-like events with durations $<30\,\mathrm{s}$ identified from consecutive pairs of 30-s LSST engineering exposures. Their occurrence rates, sky distribution, and association with tracklets indicate that the vast majority are glints from satellites and space debris. We identify 424 PSF-like isolated events with no associated tracklet.
Extending measurements of this transient foreground to the faintest magnitudes yet probed, we find an average cumulative glint rate of $(9.55 \pm 0.14)\times10^6\,\mathrm{sky}^{-1}\,\mathrm{day}^{-1}$.
Our analysis reveals a previously unreported population of glints consistent with objects in MEO. Their unresolved morphology implies characteristic durations of order $\sim$10\,ms, while their brightnesses correspond to characteristic reflector sizes of order 1--3\,mm under the assumption of perfect specular reflection.

In the future, we envisage two different paths for SMOB detections. One is to restrict the observations to the umbral region of the night sky (the anti-solar direction with a radius of about 8$^\circ$; see \citealt{Nir2021} and \citealt{Arima2026}). The primary
contaminants could be atmospheric phenomena and collisions of small bodies
(cf.\ \citealt{Arima2026}). 
In particular, a promising strategy
would be to revisit the original LSST visit cadence
of two back-to-back 15-s exposures 
for fields centered on the Earth's umbra. \
The second approach is to image the same piece of sky by 
two (or more) widely separated ground-based imagers.
Such a system cleanly separates the sources based on parallax. Specifically,
all near-earth phenomena (meteors, collisions, satellites, debris, glints) will be filtered out by the parallax 
filter.
A baseline of order 500\,km can, in principle, discriminate objects out to distances approaching 0.69\,AU. This strategy has a historical precedent: \cite{Schaefer1984} conducted simultaneous photographic observations from CTIO, Chile, and Kitt Peak, USA.  Along this line, we have started a pilot program between
ZTF (Palomar Mountain, California) and the engineering telescopes that are now being tested at the 
Sierra Remote Observatory (Fresno, California).
Ultimately, short frame time exposures with LSST and another widely separated facility 
equipped with CMOS imagers could provide the most robust path towards isolating the faint SMOBs.

%% Please use the acknowledgment and contribution environments. This will 
%% be anonomyized when the "anonymous" style option is used. 
\begin{acknowledgments}
We thank C. W. Stubbs, R. H. Lupton, E. Bellm, I. Sullivan, and J. B. Kalmbach for insightful discussions. We are also grateful to the Rubin Data Management Team, whose work made this study possible. We thank P. Antilogus, J. B. Kalmbach and M. L. Rawls for their careful review of the manuscript and constructive comments. We also thank E. Ofek and J. Tonry for helpful comments on the manuscript.

Claude and ChatGPT were used to assist with the development of plotting scripts, language editing and the automation of selected analysis tasks.
The authors take full responsibility for the scientific results.

This material is based on work supported in part by the National Science Foundation through the Cooperative Agreement AST-1258333 and the Cooperative Support Agreement AST-1202910 managed by the Association of Universities for Research in Astronomy (AURA) and the Department of Energy under Contract No. DE-AC02-76SF00515 with the SLAC National Accelerator Laboratory managed by Stanford University. Additional Rubin Observatory funding comes from private donations, grants to universities, and in-kind support from LSST-DA Institutional Members.

Lage and Tyson gratefully acknowledge support from DOE grant DE-SC0009999.

\end{acknowledgments}

\facilities{Rubin:Simonyi(LSSTCam)}

\software{Astropy \citep{astropy:2022}, 
          Matplotlib \citep{matplotlib:2007},
          LSST Science Pipelines \citep{rubinpipelines},
          Smplotlib \citep{jiaxuan_li}}

\bibliography{bib}{}
\bibliographystyle{aasjournalv7}

\appendix

\section{Chance Detections from Gaussian Noise}
\label{app:chance_detections}

It is useful to quantify the probability that a PSF-like detection arises from noise alone. This is a well-studied problem in random-field statistics \citep{Adler1981} and in astronomical source detection (\citealp{Kaiser2004}; R. H. Lupton, private communication). 
In the LSST image-differencing pipeline,
convolution of the template by the Alard--Lupton PSF-matching kernel \citep{AlardLupton1998} introduces correlations between neighboring pixels.
This covariance is subsequently removed, to first order,
by a post-subtraction decorrelation (noise-whitening) convolution,
following the approach of \citet{ReissLupton2016}, motivated by the optimal-differencing work of \citet{Zackay2016} (see also \citealt{Kaiser2004b}).
Thus, for approximately Gaussian background noise,
the decorrelated difference image can be treated to
first order as having locally white Gaussian noise.

Point-source detection then acts approximately as a matched filter: the decorrelated difference image is filtered by the PSF and local maxima are identified as candidates. The filtered image is a smooth Gaussian random field, and because PSF filtering correlates neighboring pixels, false detections must be estimated from the statistics of its peaks rather than from the single-pixel Gaussian tail. The relevant quantity is therefore the expected number of noise peaks above a given significance threshold, or SNR $\nu$.

For white noise filtered by a circular Gaussian PSF with standard deviation $w$, the expected density of noise peaks above $\nu$ is approximately
\begin{equation}
n_{\rm pk}(>\nu) \simeq \frac{1}{4\pi w^2} \frac{\nu}{\sqrt{2\pi}} e^{-\nu^2/{2}},
\label{eq:false_peaks}
\end{equation}
where $w$ is expressed in pixels \citep{Kaiser2004}. Eq.~(\ref{eq:false_peaks}) gives the peak density for a single polarity. Because a difference image can contain both positive and negative excursions, the expected number of chance detections in an image with $N_{\rm pix}$ (searchable) pixels is the following.
\begin{equation}
\langle N_{\rm false}(>\nu)\rangle
=2 N_{\rm pix}\,n_{\rm pk}(>\nu).
\label{eq:total_peaks}
\end{equation}

Adopting a representative PSF FWHM of $1\arcsec$ and the LSST pixel scale of $0\farcs2$ gives $w=\mathrm{FWHM}/2.355\simeq2.12$ pixels. For the $3.2\times10^9$ pixels in the LSST focal plane, Eq.~(\ref{eq:total_peaks}) predicts approximately $8.4\times10^2$, $7.2\times10^{-3}$, and $8.6\times10^{-14}$ chance detections per difference image above $\mathrm{SNR}=5$, 7, and 10, respectively. Thus, a single LSST difference image is expected to contain chances of order $10^3$ at $5\sigma$, whereas the Gaussian false-detection rate becomes negligible for $10\sigma$. Across our full sample of 2,955 difference images, the expected number of purely Gaussian chance detections above this threshold is only $\sim2.5\times10^{-10}$. Therefore, at our adopted threshold of $\mathrm{SNR}>10$, we expect no contribution from purely Gaussian chance detections to the candidate sample.

\section{Defocus and size of telescope}
    \label{app:Defocus}
    
\cite{loeb2024} first noted that LSST observations could be contaminated by glints from LEO satellites, while \cite{tyson2024} subsequently showed that, thanks to LSST's large aperture, LEO satellites would appear strongly out of focus in LSST images and may be faint enough to evade the alert stream. A glint appearing as an in-focus point source in LSST must therefore arise from a sufficiently distant object. This raises a natural question: how distant and how physically small must an object be to appear unresolved by LSST?

Let us derive a simple geometric estimate of their apparent angular size. For an object of physical size $S$ at distance $L$, the angular extent is approximately
\begin{equation}
    \theta_{\mathrm{obj}} \simeq \frac{S}{L}.
\end{equation}

\begin{figure}[t!]
    \centering
    \includegraphics[width=\linewidth]{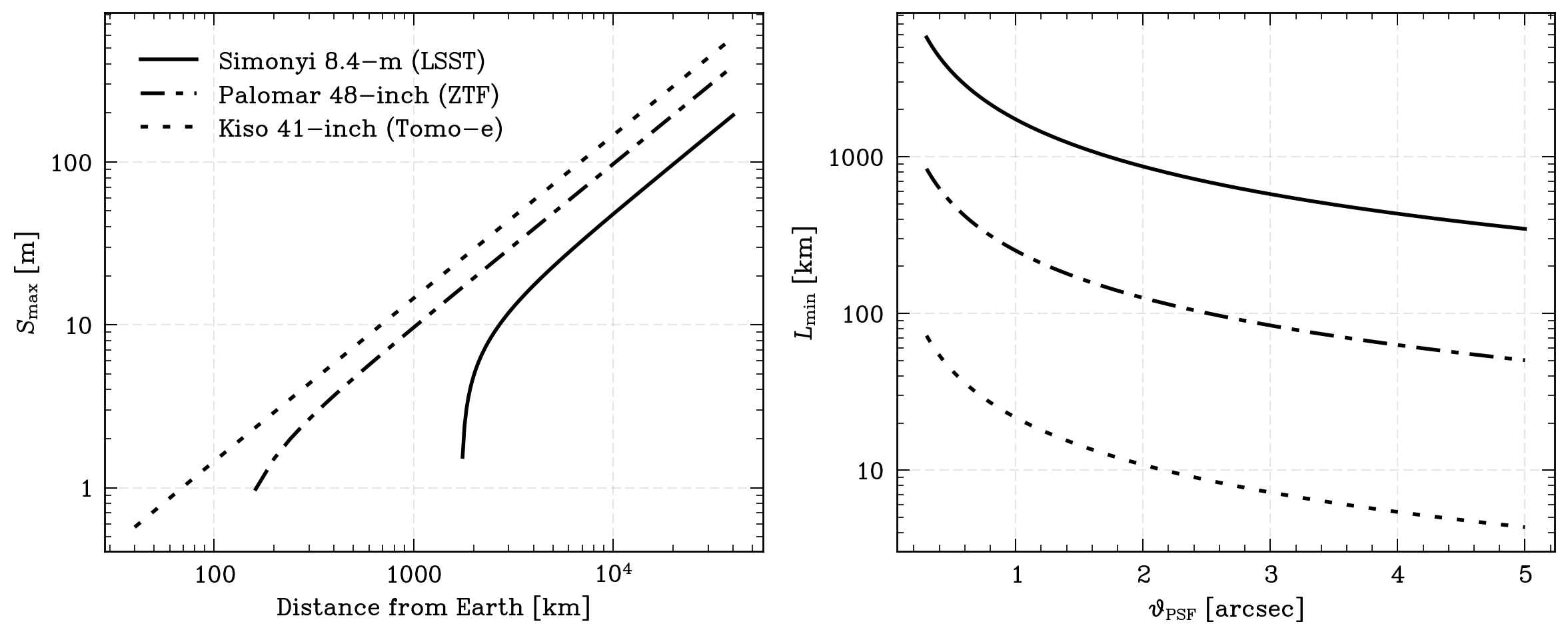}
    \caption{Defocus constraints for the Simonyi 8.4-m (LSST), Palomar 48-inch (ZTF), and Kiso 41-inch (Tomo-e Gozen) telescopes. (\textit{Left}) Maximum physical size, $S_{\rm max}$, for an object to remain unresolved as a function of distance from Earth, adopting $\theta_{\rm PSF}=1^{\prime\prime}$ for LSST, $1.2^{\prime\prime}$ for ZTF, and $3^{\prime\prime}$ for Tomo-e Gozen. (\textit{Right}) Minimum distance, $L_{\rm min}$, at which a point-like object can remain unresolved as a function of $\theta_{\rm {PSF}}$.}
    \label{fig:defocus_curve}
\end{figure}

For a telescope focused at infinity, an object at a finite distance produces a characteristic defocus blur of angular diameter
\begin{equation}
    \theta_{\mathrm{defocus}} \simeq \frac{D}{L},
\end{equation}
where $D$ is the aperture diameter of the telescope. Although the physical size of the defocus blur in the focal plane depends on the focal length $f$, conversion to an angular size on the sky introduces a factor of $1/f$, so that the defocus blur depends only on $D/L$.

For an order-of-magnitude estimate, let us assume that the intrinsic angular extent of the object and the defocus blur add in quadrature,
\begin{equation}
    \theta_{\mathrm{eff}}^2 \simeq \left(\frac{S}{L}\right)^2 + \left(\frac{D}{L}\right)^2.
\end{equation}

Requiring $\theta_{\mathrm{eff}}$ to remain below a characteristic angular resolution scale $\theta_{\mathrm{PSF}}$ then gives
\begin{equation}\label{eq:smax}
    S_{\mathrm{max}} \simeq \sqrt{L^2\theta_{\mathrm{PSF}}^2 - D^2}
\end{equation}

Taking $\theta_{\mathrm{PSF}} = 1^{\prime\prime}$ as the characteristic scale below which the glint is unresolved, we can estimate the maximum physical size of an object that could appear PSF-like as a function of distance. 

The left panel of Figure~\ref{fig:defocus_curve} shows this relation for the Simonyi 8.4-m (LSST), the Palomar 48-inch (ZTF), and the Kiso 41-inch (Tomo-e Gozen).
For LSST, the resulting curve recovers the prediction of \citet{tyson2024}: any object closer than $\sim1,732$\,km would appear out of focus, regardless of its physical size.
Thus, unresolved events in LSST images point to more distant foreground objects.

This distance threshold depends on the adopted PSF scale.
Setting the argument of the square root in Eq. (\ref{eq:smax}) to zero gives the minimum distance at which even a point-like object can appear unresolved,
\begin{equation}
    L_{\rm min} = \frac{D}{\theta_{\rm PSF}}
\end{equation}
The right panel of Figure~\ref{fig:defocus_curve} shows $L_{\rm min}$ as a function of $\theta_{\rm PSF}$ for the same three telescopes. As the PSF becomes narrower, objects must lie progressively farther away to remain unresolved; for example, improving the characteristic PSF from $1^{\prime\prime}$ to $0.8^{\prime\prime}$ increases the minimum distance by 25\%. For LSST, this shifts the unresolved-source threshold from 1,732\,km to 2,165\,km.

In practice, $L_{\rm min}$ is not a sharp cutoff, 
as somewhat closer objects may still pass our morphological selection. 
At slant ranges of 1,000 and 1,200\,km, the defocus blurs are 1\farcs73 and 1\farcs44, respectively. 
Adding these in quadrature to $1^{\prime\prime}$ seeing gives approximate image widths of $2^{\prime\prime}$ and 1\farcs75. 
Of the 4,901 events surviving our PSF-based morphological cuts, only 25 (0.5\%) have measured widths exceeding 1\farcs75, and only four exceed $2^{\prime\prime}$.
Although these approximate widths are not directly equivalent to the measured HSM widths, the compactness of our sample suggests that LEO glints are strongly suppressed.

\section{Satellite Orbits}\label{app:Satellite_Orbits}

The popular locations for satellites are Low-Earth Orbits (LEO; ($<1,000\,$km to $2,000\,$km -- the lower edge of the van Allen belt) and 
GEO-synchronous (GEO; 35,876\,km, nominal). LEO is popular with communication
companies (proximity to customers, low $\Delta v$), earth scientists
and astronomers (sun-synchronous, protected from radiation belt) while
GEO is used primarily for telecommunications and broadcasting, earth and weather monitoring, satellite navigation enhancement, and defense. 
Orbital radius  between that of LEO and GEO is primarily occupied by the well
known global navigation satellite systems\footnote{India's  
Regional navigation systems, NAVIC, and Japan's, QZSS,
are in inclined GEO orbits and are designed to provide localization services to a specific region.} such as GPS, GLONASS, Galileo and
BeiDou.
The number of satellites in LEO stands currently (October 2026) at $\sim$17,000 active satellites (and increasing rapidly; \citealt{Mcdowell2026}). 

GEO currently has 569 active satellites. 
Due to the equatorial ellipticity of the Earth, a GEO satellite drifts in longitude (towards 
$73.5^\circ$E or $105^\circ$W) over months to years
\citep{bbk+62}.
The perturbation due to the moon and sun led to 
a growth of orbital inclination $i$, to
$i=\pm 15^\circ$ with a cycle of 53\,years \citep{ac64}. In addition, owing to the influence of moon,
the orbit starts to acquire eccentricity. During the operational phase, the satellites are kept in the assigned orbital phase by 
onboard propulsion systems.
At the end of their operational lifetime,
the modern practice is for the operators to
increase the satellite to a higher altitude (300\,km to 400\,km) -- the so-called
``graveyard''. These satellites, lacking propulsion, are subject to the perturbations
discussed above and acquire inclined orbits with some eccentricity. 
In due course, the orbits can become
chaotic \citep{gdg+16}.
The graveyard is thus a ring with 
a declination range\footnote{Note that owing to parallax, the topocentric declination
of a GEO satellite depends on the location of the observer; Observers away from
the equator would see an offset of up to 8$^\circ$.}
of $\pm 15^\circ$. The European Space Agency (ESA)
manages a database of space debris -- the DISCOS
database. The agency  formally defines the graveyard by a declination of $\pm 25^\circ$ 
and an altitude of 35,586--35,986\,km. DISCO lists about 5,000 objects
in the graveyard.

During the period 2001--2004 ESA's optical GEO program (based 
on the 1-m Space Debris Telescope on Tenerife Island) discovered high declination
faint and variable objects in GEO. These are now referred to as HAMR 
(high area-to-mass ratio) objects. The new ingredient is solar radiation pressure, which becomes important for objects with area-to-mass ratios in the range of 1 to 50
m$^2$/kg (see \citealt{gdg+16} for the underlying theory).  
A comprehensive summary of the optical photometry of a sample
of HAMR objects (to about 20\,mag) can be found in \cite{smf06}.  

\section{Catalog and Host-Association Analysis of the Isolated Sample}\label{app:isolated_catalog}

Here, we present a detailed summary of the isolated sample of 424-events introduced in \S\ref{sec:results}. Table~\ref{tab:isolated_short} presents the subset with the most significant apparent galaxy associations; the complete table is available in machine-readable form. For each event, we report its magnitude, closest cataloged galaxy, angular separation from that galaxy, and the corresponding probability of chance coincidence $P_{\rm chance}$. We compute $P_{\rm chance}$ following \citet{Bloom2002}, using measurements of the cumulative surface density of galaxies as a function of apparent magnitude, taken from the number counts of \citet{Hogg1997}. Galaxies are drawn from the 4-m Blanco telescope Legacy Survey DR10 \citep{Dey2019}.

\begin{figure*}
    \centering
    \includegraphics[width=0.9\linewidth]{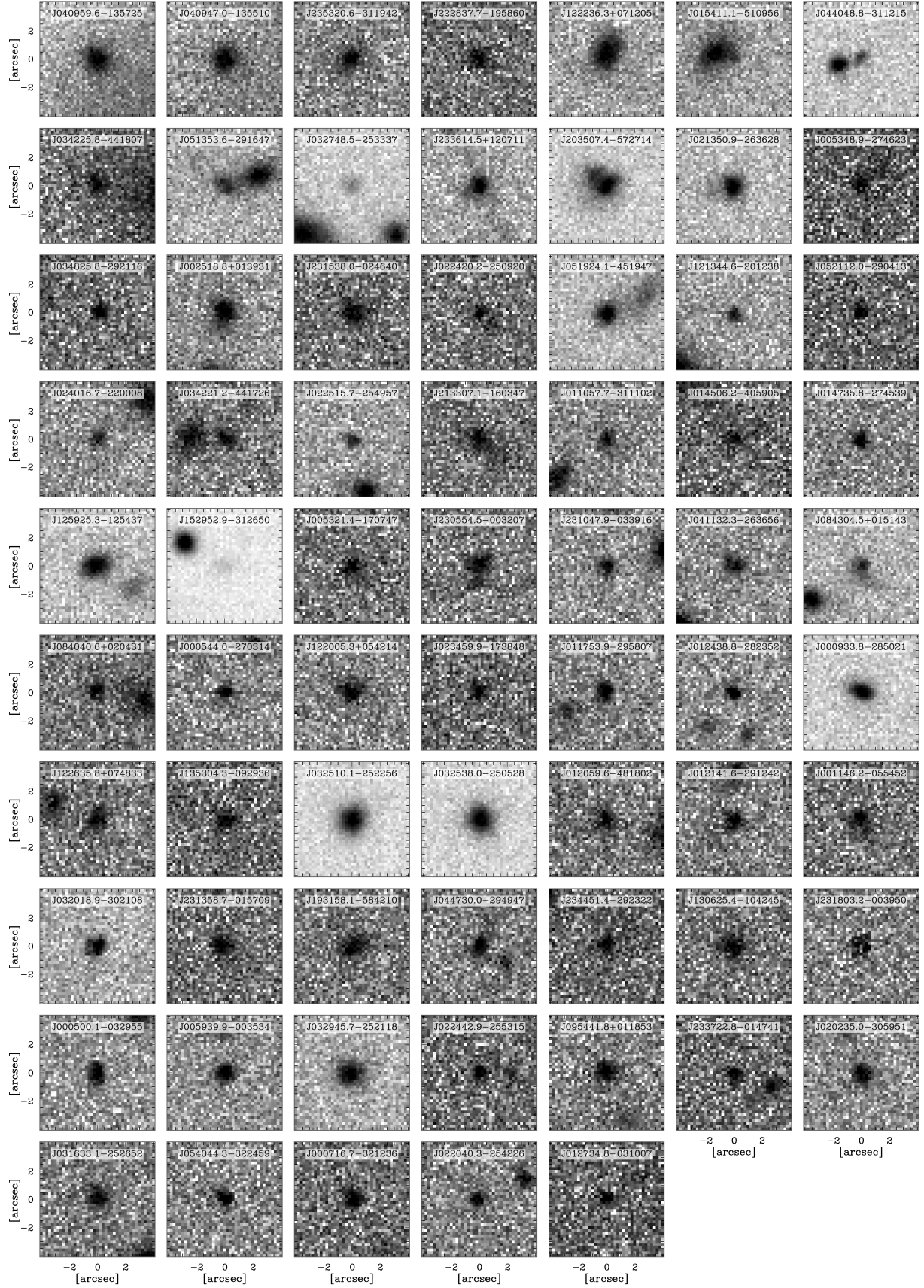}
    \caption{Image cutouts of candidates from the conservative 424-event isolated sample with $P_{\rm chance}<0.05$ relative to the nearest cataloged galaxy. These events have the lowest chance-coincidence probabilities in the sample, but positional coincidence alone does not establish a physical association with the galaxy. A foreground origin therefore remains plausible.}
    \label{fig:isolated_cutouts}
\end{figure*}

These galaxy associations are reported to assess whether any candidate has a potentially significant nearby host but should not by themselves be interpreted as evidence of a physical association. Only 114 of the 424 events (26.9\%) have $P_{\rm chance}<0.1$, while the remaining 73.1\% have associations readily attributable to chance projection. For completeness, Figure~\ref{fig:isolated_cutouts} shows image cutouts for the subset with $P_{\rm chance}<0.05$, for which an apparent association with a nearby galaxy is less likely to arise by chance. Even for these candidates, morphology and positional coincidence alone cannot establish an astrophysical origin; foreground objects remain a plausible explanation, and simultaneous two-station imaging would be required to distinguish the two possibilities robustly.

\section{Orbital Constraints from Tracklets}
	\label{app:tracklet}

\begin{figure*}
    \centering
    \includegraphics[width=\linewidth]{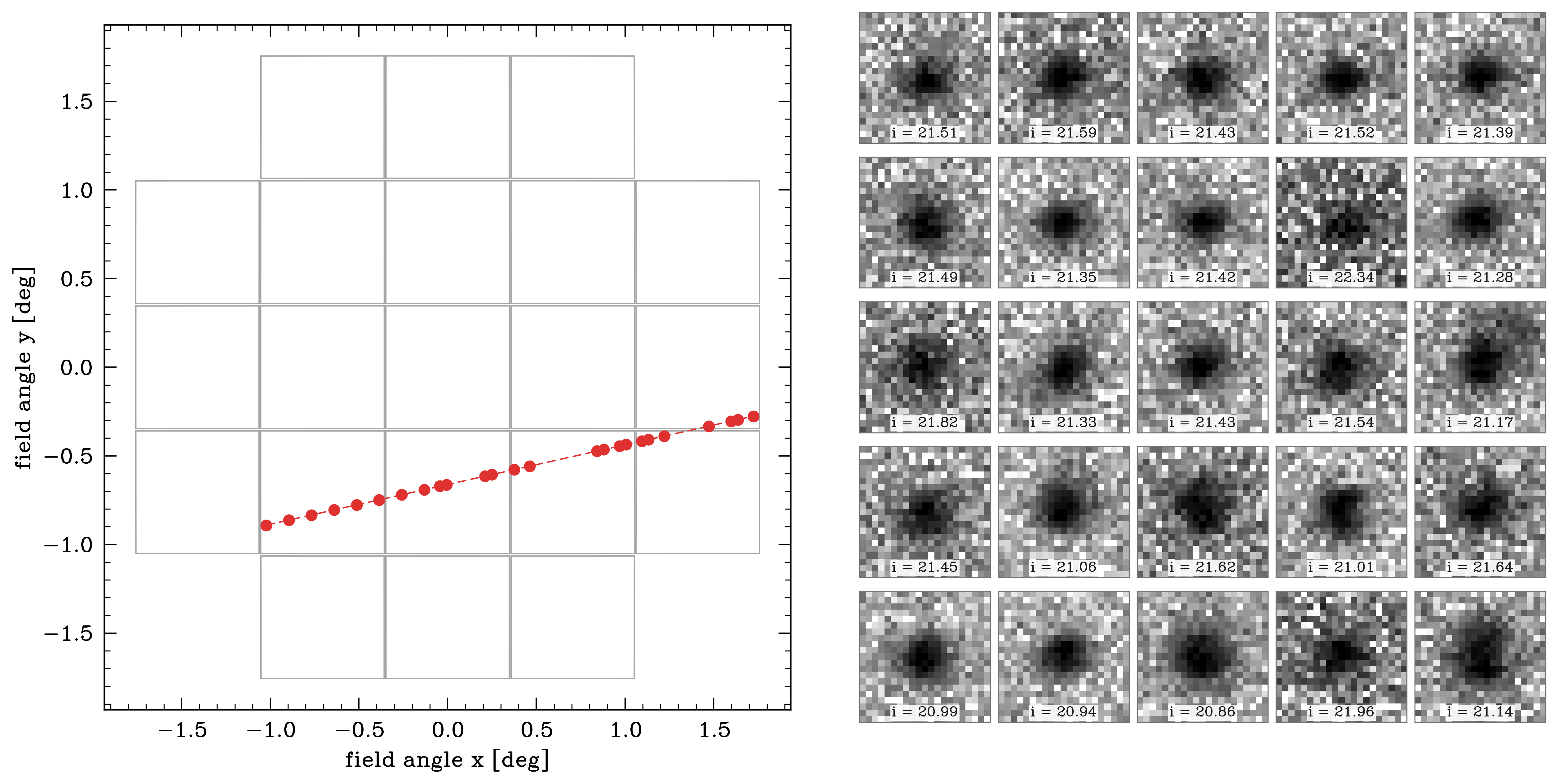}
    \caption{(\textit{Left}) Tracklet footprint in the pair of exposures where it was detected. The tracklet spans 2.816$^\circ$, corresponding to an angular rate of $151^{\prime\prime}\,\mathrm{s}^{-1}$ and an upper limit on the orbital altitude of 7,269\,km. (\textit{Right}) All the cutouts from the tracklet at about 21\,mag showing point-source like features despite showing some structure as well. 
    The glints do not appear in sequential order along the tracklet,
    raising the possibility that they originate from multiple fragments traveling along similar orbits rather than from a single tumbling object.
    In this case, the inferred orbital-altitude limit would not apply.
    }
    \label{fig:tracklet}
\end{figure*}

In \S\ref{app:Defocus}, we showed that objects in low-Earth orbit (LEO) cannot produce unresolved in-focus signatures in LSST images. We can test this argument empirically using the tracklets identified in our sample. For example, the tracklet shown in Figure \ref{fig:tracklet} spans 2.816$^\circ$ across the focal plane over the two consecutive 30-s exposures (total time of 67\,s, accounting for readout), corresponding to a minimum angular rate of 0.0418$^\circ\,\mathrm{s}^{-1}$ (151$^{\prime\prime}\,\mathrm{s}^{-1}$). Individual glints along the tracklet are separated by a median of 2.2\,s. Because the tracklet may extend beyond the focal plane, this angular rate is a lower limit.
The glints do not appear in sequential order along the tracklet, suggesting that they could originate from multiple fragments that have drifted apart along similar orbits \citep{Johnson2009, Shu2026}, rather than from a single tumbling object.

Assuming the object is in a circular orbit, the apparent angular rate can be approximated as the transverse orbital velocity divided by the slant range,
\begin{equation}
\omega \simeq \frac{v_{\rm orb}}{d},
\end{equation}
where
\begin{equation}
v_{\rm orb}(h) = \sqrt{\frac{GM_\oplus}{R_\oplus+h}}
\end{equation}
is the circular-orbit velocity at altitude $h$, and $d$ is the line-of-sight distance from the observatory to the object (slant range). For an observation at zenith angle $z$, the slant range is given by
\begin{equation}
d(h,z) =
\sqrt{(R_\oplus\cos z)^2 + 2R_\oplus h + h^2} - R_\oplus\cos z
\end{equation}
a standard result from spherical geometry (see, e.g., \citealt{Vasylyev2019}). This conversion assumes a circular orbit. For an eccentric orbit, the instantaneous orbital velocity also depends on orbital phase, so a measured angular rate does not correspond to a unique altitude.

Using the estimated angular rate derived from the observed tracklet span, we solve for the orbital altitude $h$ through
\begin{equation}\label{eq:altitude}
\omega_{\rm obs}
=
\frac{1}{d(h,z)}
\sqrt{\frac{GM_\oplus}{R_\oplus+h}}.
\end{equation}
where we have ignored the rotation of the Earth.

For the tracklet presented in Figure \ref{fig:tracklet}, we obtain a maximum orbital altitude of 7,269\,km and a slant range of 7,399\,km. The corresponding orbital velocity is $5.4\,\mathrm{km}\,\mathrm{s}^{-1}$, with an orbital period of 264\,min. This upper-limit altitude places the object in MEO and is consistent with the distance constraints derived from the defocus calculation in Figure~\ref{fig:defocus_curve}. At this distance, the requirement that the source is unresolved constrains its characteristic size to $\leq 30\,\mathrm{m}$, including objects ranging from meters-scale structures down to millimiter-scale debris. 

Having constrained the orbital altitude, we estimate the derbis radius $r_{\rm debris}$ required to reproduce the brightest glint in the tracklet \citep[cf.][]{Nir2021}. For the measured angular rate of $(151^{\prime\prime}\,{\rm s}^{-1})$, a glint must last no longer than
\begin{equation}
t_{\rm g} \lesssim \frac{1^{\prime\prime}}{151^{\prime\prime}\,{\rm s}^{-1}}
\simeq 6.6\,{\rm ms}
\end{equation}
for its motion during the glint to remain below $\theta_{\rm PSF} \leq 1^{\prime\prime}$ and hence appear unresolved. This is substantially shorter than the $t_{\rm g}\sim0.1-0.5\,{\rm s}$ durations measured for glints from GEO objects \citep{Karpov2016,Nir2021}. Because the measured magnitude is diluted over the 30-s exposure, the corresponding instantaneous glint magnitude is
\begin{equation}
    m_{\rm g} = m_{\rm obs} - 2.5\log_{10}(t_{\rm exp}/t_{\rm g})
    \simeq m_{\rm obs} - 9.14 .
\end{equation}

For a specularly reflecting flat surface, the reflected-to-solar flux ratio is
\begin{equation}
\frac{F_{\rm g}}{F_\odot}
=10^{-0.4(m_{\rm g}-m_\odot)}
=\rho\left(\frac{r_{\rm debris}}
{d\,\theta_\odot}\right)^2 ,
\end{equation}
where $\rho$ is the albedo, $d$ is the slant range, and $\theta_\odot\simeq4.65\times10^{-3}\,{\rm rad}$ is the angular radius of the Sun. Solving for $r_{\rm debris}$ gives
\begin{equation}
r_{\rm debris}
= \frac{d\,\theta_\odot}{\rho^{1/2}}\,
10^{(m_\odot-m_{\rm g})/5}
\end{equation}

Adopting $\rho=0.175$, the conventional albedo used in optical size estimates of orbital debris \citep{Africano2004}, together with $d\simeq7{,}399\,{\rm km}$, and $m_{\rm obs}\simeq21$\,mag (the brightest glint in Figure \ref{fig:tracklet}),  gives $r_{\rm debris}\sim1\,{\rm mm}$ for the representative tracklet in Figure \ref{fig:tracklet}.
Since
$r_{\rm debris} \propto \rho^{-1/2}$, 
a more reflective surface would imply an even smaller object.

These estimates represent lower limits at the assumed
orbital altitudes, since shorter glint durations or
less efficient reflection would require larger surfaces.
For comparison, \citet{tanaka2025} show that a diffuse 
Lambertian reflector observed at a phase angle of $90^\circ$ 
would require a size larger by a factor of $\sim500$ 
for the same brightness. 
Thus, the $\sim1$--$10\,{\rm mm}$ sizes inferred under the specular assumption would correspond to roughly 
$\sim0.5$--$5\,{\rm m}$ under diffuse reflection. 
Aluminized multilayer insulation (MLI), however, is predominantly specular \citep{Fruh2012}.

We apply these orbital-altitude and size calculations to all identified tracklets to obtain the distributions shown in Figure~\ref{fig:tracklet_hist}.
The resulting upper limits on the glint duration have a median of $8\,\mathrm{ms}$ and span $4$--$19\,\mathrm{ms}$ across the tracklet sample.

\begin{deluxetable*}{lccccclcccc}
\tabletypesize{\scriptsize}
\tablewidth{0pt}
\tablecaption{Isolated candidates with $P_{\rm chance} < 0.05$. $\theta$ is the separation from the galaxy minimising $P_{\rm chance}$. The full table is available in machine-readable format in the online material.\label{tab:isolated_short}}
\tablehead{\colhead{Name} & \colhead{Visit} & \colhead{$m_{\rm AB}$} & \colhead{$\theta$} & \colhead{$P_{\rm chance}$} & \colhead{} & \colhead{Name} & \colhead{Visit} & \colhead{$m_{\rm AB}$} & \colhead{$\theta$} & \colhead{$P_{\rm chance}$}}
\startdata
LSST-FT J040959.6-135725 & 2025121900029 & 20.34 & 0.06 & 0.0000 &  & LSST-FT J084304.5+015143 & 2025122200339 & 22.53 & 4.30 & 0.0263 \\
LSST-FT J040947.0-135510 & 2025121900029 & 20.72 & 0.06 & 0.0000 &  & LSST-FT J084040.6+020431 & 2025122200336 & 23.02 & 3.46 & 0.0263 \\
LSST-FT J235320.6-311942 & 2025120800013 & 22.61 & 0.34 & 0.0003 &  & LSST-FT J000544.0-270314 & 2025120300037 & 22.21 & 4.02 & 0.0269 \\
LSST-FT J222837.7-195860 & 2025082500336 & 22.99 & 0.30 & 0.0011 &  & LSST-FT J122005.3+054214 & 2025052200088 & 22.10 & 1.44 & 0.0271 \\
LSST-FT J122236.3+071205 & 2025052200131 & 22.06 & 0.89 & 0.0012 &  & LSST-FT J023459.9-173848 & 2025112900014 & 22.01 & 28.86 & 0.0278 \\
LSST-FT J015411.1-510956 & 2025112800019 & 22.56 & 1.38 & 0.0015 &  & LSST-FT J011753.9-295807 & 2025102700152 & 22.67 & 3.10 & 0.0278 \\
LSST-FT J044048.8-311215 & 2025121700230 & 22.29 & 1.65 & 0.0042 &  & LSST-FT J012438.8-282352 & 2025111400055 & 22.08 & 5.95 & 0.0285 \\
LSST-FT J034225.8-441807 & 2025121100097 & 23.03 & 7.19 & 0.0057 &  & LSST-FT J000933.8-285021 & 2025102700200 & 20.58 & 17.37 & 0.0292 \\
LSST-FT J051353.6-291647 & 2025111200263 & 21.67 & 2.46 & 0.0058 &  & LSST-FT J122635.8+074833 & 2025052200120 & 22.51 & 3.23 & 0.0294 \\
LSST-FT J032748.5-253337 & 2025111500199 & 22.24 & 5.01 & 0.0060 &  & LSST-FT J135304.3-092936 & 2025062000151 & 23.09 & 0.78 & 0.0303 \\
LSST-FT J233614.5+120711 & 2025110700043 & 21.31 & 0.99 & 0.0070 &  & LSST-FT J032510.1-252256 & 2025112000066 & 20.57 & 1.50 & 0.0304 \\
LSST-FT J203507.4-572714 & 2025061900435 & 19.66 & 1.34 & 0.0082 &  & LSST-FT J032538.0-250528 & 2025112000066 & 20.72 & 1.11 & 0.0309 \\
LSST-FT J021350.9-263628 & 2025112200026 & 20.74 & 1.26 & 0.0086 &  & LSST-FT J012059.6-481802 & 2025112200165 & 23.20 & 4.49 & 0.0311 \\
LSST-FT J005348.9-274623 & 2025111400026 & 22.71 & 1.67 & 0.0092 &  & LSST-FT J012141.6-291242 & 2025102700148 & 22.88 & 2.19 & 0.0314 \\
LSST-FT J034825.8-292116 & 2025121700206 & 22.69 & 11.00 & 0.0094 &  & LSST-FT J001146.2-055452 & 2025112100024 & 22.34 & 16.40 & 0.0337 \\
LSST-FT J002518.8+013931 & 2025091500263 & 21.34 & 0.61 & 0.0095 &  & LSST-FT J032018.9-302108 & 2025121700087 & 21.87 & 8.08 & 0.0361 \\
LSST-FT J231538.0-024640 & 2025111500083 & 21.84 & 1.15 & 0.0097 &  & LSST-FT J231358.7-015709 & 2025110900025 & 22.18 & 1.70 & 0.0364 \\
LSST-FT J022420.2-250920 & 2025112400019 & 22.94 & 1.02 & 0.0097 &  & LSST-FT J193158.1-584210 & 2025090200026 & 21.53 & 1.92 & 0.0369 \\
LSST-FT J051924.1-451947 & 2025120900307 & 21.93 & 2.97 & 0.0106 &  & LSST-FT J044730.0-294947 & 2025121700236 & 22.36 & 2.25 & 0.0369 \\
LSST-FT J121344.6-201238 & 2025052400228 & 22.98 & 7.17 & 0.0117 &  & LSST-FT J234451.4-292322 & 2025102700189 & 22.63 & 2.13 & 0.0377 \\
LSST-FT J052112.0-290413 & 2025111200281 & 22.72 & 17.64 & 0.0118 &  & LSST-FT J130625.4-104245 & 2025052400372 & 22.20 & 7.30 & 0.0398 \\
LSST-FT J024016.7-220008 & 2025121700266 & 22.20 & 5.49 & 0.0119 &  & LSST-FT J231803.2-003950 & 2025110900037 & 22.47 & 10.01 & 0.0399 \\
LSST-FT J034221.2-441726 & 2025121100097 & 22.74 & 2.37 & 0.0122 &  & LSST-FT J000500.1-032955 & 2025121700050 & 21.63 & 5.27 & 0.0419 \\
LSST-FT J022515.7-254957 & 2025112400013 & 22.14 & 3.75 & 0.0136 &  & LSST-FT J005939.9-003534 & 2025110900133 & 22.21 & 2.80 & 0.0422 \\
LSST-FT J213307.1-160347 & 2025102200113 & 22.09 & 1.65 & 0.0148 &  & LSST-FT J032945.7-252118 & 2025112000067 & 22.04 & 1.50 & 0.0425 \\
LSST-FT J011057.7-311102 & 2025102700138 & 22.50 & 1.41 & 0.0171 &  & LSST-FT J022442.9-255315 & 2025112400019 & 22.68 & 2.24 & 0.0428 \\
LSST-FT J014506.2-405905 & 2025120200017 & 21.83 & 2.27 & 0.0172 &  & LSST-FT J095441.8+011853 & 2025052000099 & 22.95 & 3.96 & 0.0431 \\
LSST-FT J014735.8-274539 & 2025121700155 & 22.27 & 8.51 & 0.0189 &  & LSST-FT J233722.8-014741 & 2025110800045 & 22.66 & 2.85 & 0.0435 \\
LSST-FT J125925.3-125437 & 2025052400309 & 21.05 & 3.06 & 0.0203 &  & LSST-FT J020235.0-305951 & 2025111900043 & 21.75 & 2.98 & 0.0437 \\
LSST-FT J152952.9-312650 & 2025072800085 & 22.30 & 8.85 & 0.0220 &  & LSST-FT J031633.1-252652 & 2025111500180 & 21.99 & 5.77 & 0.0456 \\
LSST-FT J005321.4-170747 & 2025112500034 & 22.49 & 1.36 & 0.0223 &  & LSST-FT J054044.3-322459 & 2025121700216 & 22.34 & 12.83 & 0.0463 \\
LSST-FT J230554.5-003207 & 2025110900011 & 21.57 & 22.78 & 0.0239 &  & LSST-FT J000716.7-321236 & 2025120800013 & 22.48 & 7.49 & 0.0466 \\
LSST-FT J231047.9-033916 & 2025102200195 & 22.16 & 4.30 & 0.0244 &  & LSST-FT J022040.3-254226 & 2025112200030 & 22.27 & 3.69 & 0.0476 \\
LSST-FT J041132.3-263656 & 2025111000205 & 22.20 & 6.08 & 0.0252 &  & LSST-FT J012734.8-031007 & 2025112700013 & 22.42 & 3.16 & 0.0492 \\
\enddata
\tablecomments{Names encode the J2000 position. $\theta$ is the separation from the galaxy minimising $P_{\rm chance}$, and Visit is the exposure the source was detected in. Bandpass, signal-to-noise, host magnitude and type are given in the machine-readable table.}
\end{deluxetable*}

%\bibliography{PASPsample701}{}
\end{document}